\documentclass{emulateapj}

\usepackage{apjfonts}
\usepackage{aas_macros}
\usepackage{subfigure}

\usepackage{xcolor}

\usepackage[flushleft]{threeparttable}

\begin{document}

\title{A unified binary neutron star merger magnetar model for the Chandra X-ray transients CDF-S XT1 and XT2 }

\author{Hui Sun\altaffilmark{1},Ye Li \altaffilmark{2}, Bin-Bin Zhang\altaffilmark{3,4}, Bing Zhang\altaffilmark{5}, Franz E. Bauer\altaffilmark{6,7,8}, Yongquan Xue\altaffilmark{9,10}, Weimin Yuan\altaffilmark{1,11}}
\altaffiltext{1}{Key Laboratory of Space Astronomy and Technology, National Astronomical Observatories, Chinese Academy of Sciences, Beijing 100012, China, hsun@nao.cas.cn;}

\altaffiltext{2}{Kavli Institute for Astronomy and Astrophysics, Peking University, Beijing 100871, China;}

\altaffiltext{3}{School of Astronomy and Space Science, Nanjing University, Nanjing 210093, China;}
\altaffiltext{4}{Key Laboratory of Modern Astronomy and Astrophysics (Nanjing University), Ministry of Education, China}

\altaffiltext{5}{Department of Physics and Astronomy, University of Nevada, Las Vegas, NV 89154, USA, zhang@physics.unlv.edu;}

\altaffiltext{6}{Instituto de Astrof{\'{\i}}sica and Centro de Astroingenier{\'{\i}}a, Facultad de F{\'{i}}sica, Pontificia Universidad Cat{\'{o}}lica de Chile, Casilla 306, Santiago 22, Chile;}
\altaffiltext{7}{Millennium Institute of Astrophysics (MAS), Nuncio Monse{\~{n}}or S{\'{o}}tero Sanz 100, Providencia, Santiago, Chile;}
\altaffiltext{8}{Space Science Institute, 4750 Walnut Street, Suite 205, Boulder, CO 80301;}
\altaffiltext{9}{CAS Key Laboratory for Research in Galaxies and Cosmology, Department of Astronomy, University of Science and Technology of China, Hefei, 230026, China;}
\altaffiltext{10}{School of Astronomy and Space Science, University of Science and Technology of China, Hefei, 230026, China;}

\altaffiltext{11}{School of Astronomy and Space Science, University of Chinese Academy of Sciences, 19A Yuquan Road, Beijing, 100049, China}
\begin{abstract}

Two bright X-ray transients were reported from the Chandra Deep Field South (CDF-S) archival data, namely CDF-S XT1 and XT2. Whereas the nature of the former is not identified, the latter was suggested as an excellent candidate for a rapidly spinning magnetar born from a binary neutron star (BNS) merger. Here we propose a unified model to interpret both transients within the framework of the BNS merger magnetar model. According to our picture, CDF-S XT2 is observed from the ``free zone'' where the magnetar spindown powered X-ray emission escapes freely, whereas CDF-S XT1 originates from the ``trapped zone'' where the X-ray emission is initially blocked by the dynamical ejecta and becomes transparent after the ejecta is pushed to a distance where Thomson optical depth drops below unity. 
We fit the magnetar model to the light curves of both transients and derived consistent parameters for the two events, with magnetic field, initial spin period and X-ray emission efficiency being ($B_p=10^{16}\,G$, $P_i=1.2\,\rm ms$, $\eta = 0.001$) and ($B_p=10^{15.8}\,G$, $P_i=4.4\, \rm ms$, $\eta = 0.001$) for XT1 and XT2, respectively. The ``isotropic equivalent'' ejecta mass of XT1 is $M_{\rm ej} \sim 10^{-3}$ $M_{\odot}$, while it is not constrained for XT2. Our results suggest that more extreme magnetar parameters are required to have XT1 detected from the trapped zone.  The model parameters for both events are generally consistent with those derived from SGRB X-ray plateau observations. 
The host-galaxy properties of both transients are also consistent with those of SGRBs. The event rate densities of both XT1 and XT2 are consistent with that of BNS mergers.
\end{abstract}

\section{Introduction}
The discovery of the first gravitational-wave event GW170817 from binary neutron star (BNS) merger \citep{abbott2017a} and its broadband electromagnetic (EM) counterparts \citep{abbott2017b} ushered in the multimessenger era of astrophysics. The EM counterparts, including the short gamma-ray burst (SGRB) 170817A, kilonova AT2017gfo, and broadband afterglow, have confirmed the related theoretical models and posed interesting constraints on model parameters (e.g., see \citealt{metzger2017} for a summary). The merger product of this event is not well constrained. Several groups argued that the merger product is a black hole (BH), probably formed after a brief hypermassive neutron star (HMNS) phase \citep[e.g.][]{margalitmetzger2017, pooley2018, rezzolla2018, ruiz2018}. However, the possibility of a long-lived supramassive neutron star (SMNS) or stable neutron star (SNS) with a low dipolar magnetic field cannot be ruled out from the data \citep{ai2018}, and the existence of such a long-lived remnant is helpful to interpret some of the observations \citep{yu2018,li2018,piro2019}.

The possibility of BNS mergers producing a long-lived neutron star (NS) has been suggested in the literature to interpret some of the X-ray features in SGRB afterglows, including X-ray flares \citep{dai2006,gao2006}, extended emission \citep{metzger2008}, and especially the so-called internal plateau observed in a good fraction of SGRBs (\citealt{rowlinson2010}, \citealt{ rowlinson2013}, hereafter R13, \citealt{lv2015}, hereafter L15, \citealt{gao2016}, \citealt{li2016a,li2017}, cf. \citealt{rezzolla2015}). The existence of such long-lived neutron star requires that the equation of state of the neutron star is relatively stiff, which is indeed supported by some recent observations of Galactic massive neutron stars \citep{lattimerprakash2010,cromatie2019}. The X-ray plateau can be powered by the dissipation of a magnetar wind which is essentially isotropic for a rapidly spinning magnetar. In such a case, one would expect BNS-merger powered X-ray transients with no SGRB detection if the line of sight misses the bright jet zone of the event \citep{zhang2013,sun2017}.

\cite{sun2017} studied the possible X-ray light curves of a long-lived magnetar generated from BNS mergers for different observer's viewing angles. In particular, they defined three geometric zones where the observer would view different light curves:
\begin{itemize}
\item The jet zone. In this zone, the observer would see a bright SGRB. SGRBs with extended emission or internal plateau belong to this configuration;
\item The free zone. In this zone, the observer can see X-rays freely but not $\gamma$-rays. There could still be a GRB 170817A-like weak GRB in that viewing direction which clears a funnel to allow X-rays to escape, but such a SGRB is not detectable if the distance is large enough, e.g. $\gtrsim 80$ Mpc in the case of GRB 170817A \citep{zhang2018};
\item The trapped zone. In this zone, X-rays are initially trapped behind the ejecta from the BNS merger system, but eventually become free when the ejecta is pushed to a large enough radius. 
\end{itemize}
Assuming that the X-ray radiation efficiency does not sensitively depend on the spindown luminosity, the X-ray light curve in the jet/free zone would generally follow the dipole spindown law, and thus appears as a bright plateau followed by a decay with a temporal index between -1 and -2 if the magnetar does not collapse, or steeper than -3 if it collapses. In the trapped zone, the X-ray emission is initially absorbed until the ejecta reaches the transparent radius, i.e. when the photosphere radius has traversed the ejecta from larger radii to smaller radii. The X-ray light curve, therefore, should show a rapid rise before the transparent time (as the optical depth $\tau$ drops below unity) and then follow the dipolar radiation law afterward. For typical magnetar parameters, the spindown timescale is shorter than the transparent time. As a result, the light curves in the trapped zone should follow a decay segment after the rising phase. The transparent time depends on the surface magnetic field and the initial spin period of the magnetar, and the opacity and the mass of the ejecta. Through Monte Carlo simulations, \cite{sun2017} found that for typical parameters of magnetars and ejecta, the peak luminosity of the X-ray counterparts of BNS mergers should be around $10^{46.4}$ and $10^{49.6}$ $\rm erg\,s^{-1}$ for the line of sight in the trapped zone and free zone, respectively. One important question is whether such events have been detected by current telescopes and how their event rate density compares with those of other X-ray transients in the universe \citep{sun2015}.

The 7 Ms Chandra Deep Field South (CDF-S) archive data \citep{luo2017} is an excellent resource to search for such transients \citep{zheng2017, yang2019}. Recently, a peculiar X-ray transient, CDF-S XT2, was discovered to be associated with a host galaxy at $z \sim 0.738$ \citep{xue2019}. Its light curve tracks well the prediction of the spindown luminosity evolution of a millisecond magnetar, with a peak rest-frame 0.3-10 $\rm keV$ luminosity of $L_{\rm peak,XT2} = 2.7^{+6.3}_{-2.3} \times 10^{45}$ $\rm erg\,s^{-1}$. The light curve shows a shallow plateau phase lasting for about 2 ks followed by a $\sim t^{-2}$ decay. The source was located in the outskirts of a host galaxy with a low star-formation rate. Both the type of the host galaxy and its offset from the center of the galaxy are typical for known SGRBs \citep{xue2019}. All the data strongly suggest that XT2 originates from a rapidly spinning magnetar formed from a BNS merger.

Besides XT2, there was another bright transient, CDF-S XT1, also discovered in the Chandra Deep Field South survey \citep{bauer2017}. The light curve shows a quick rise at around 110 s to a peak flux (0.3-10 $\rm keV$) of $5.1 \times 10^{-12}$ $\rm erg\,s^{-1}$ followed by a power-law decay with a slope $ -1.53\pm 0.27 $ \citep{bauer2017}. The average spectral slope is $\Gamma = 1.43 ^{+0.23}_{-0.13}$. No additional X-rays above the background rate have been detected in coincidence with this position from Chandra and XMM Newton archives. 
The photometric redshift of the host galaxy is $z_{\rm ph}=2.23$ (0.39-3.21 at 2$\sigma$ confidence), which leads to a peak luminosity (2-10 $\rm keV$) $6.8 \times 10^{46}$ $\rm erg\,s^{-1}$ ($\sim 1-140 \times 10^{45}$ $\rm erg\,s^{-1}$ over the redshift range). The faint host galaxy ($m_{R}=27.5$ $\rm mag$) was identified from the CANDELs survey. \cite{bauer2017} discussed several possible physical origins of this transient but no conclusive result was claimed. The BNS merger magnetar model was considered as one of the possible scenarios to interpret the data, but other possibilities including a tidal disruption event, a supernova shock breakout and a GRB orphan afterglow cannot be completely ruled out, even though they are disfavored for various reasons.

Motivated by the discovery of XT2, in this paper we reinvestigate XT1 and propose a unified model to interpret both transients within the framework of a BNS merger magnetar model. 
The detailed description of the model and the light curve fitting are presented in Sections 2 and 3, respectively. In Sections 4 and 5, we compare the properties of these two transients with SGRBs in terms of the X-ray plateau and host properties and estimate the event rate density of these transients. The results are summarized in Section 6.

\section{Model}
\subsection{Magnetar wind emission}
Consider a rapidly spinning magnetar produced from a BNS merger. It loses its angular momentum through both magnetic dipole radiation and quadrupole gravitational wave (GW) radiation, with the energy loss rate described as \citep{shapiroteukolsky1983,zhangmeszaros2001}
\begin{equation}
\dot{E}=I\Omega \dot{\Omega}=-\frac{B_p^2R^6\Omega^4}{6c^3}-\frac{32GI^2 \epsilon ^2 \Omega^6}{5c^5},
\label{eq:Edot}
\end{equation}
where $\Omega=2\pi/P$ is the angular frequency and $\dot{\Omega}$ its time derivative, $I$ is the moment of inertia, $B_p$ is the dipolar field strength at the magnetic poles on the NS surface, $R$ is the radius of the NS, and $\epsilon$ is the ellipticity of the NS. The first term is the magnetic dipole radiation or magnetar wind spindown term, and the second term describes the GW radiation energy loss rate. 

Similar to \cite{sun2017}, we parameterize the X-ray emission luminosity due to magnetar wind dissipation as being proportional to the dipole spindown luminosity, i.e.,
\begin{equation}
L_{\rm X,jet/free}(t)=\eta L_{\rm sd} \equiv \eta \frac{B_p^2R^6\Omega^4}{6c^3},
\label{eq:X-free}
\end{equation}
where $\eta$ is the efficiency of converting the dipole spindown luminosity to the observed X-ray luminosity. By solving the $\Omega$-evolution using Eq.(\ref{eq:Edot}) and plugging it into Eq.(\ref{eq:X-free}), one can obtain the X-ray light curve given a quantified $\eta$ evolution. For simplicity, we take $\eta$ as a constant. 

In the jet zone or free zone, the light curve starts with a relatively flat plateau, which is followed by a decay of $\propto t^{-2}$ (dipole spindown dominated) or $\propto t^{-1}$ (GW spindown dominated). 

In the trapped zone, the X-ray photons are initially trapped in the ejecta and energize the ``merger-nova'' (which is the kilonova with extra energy injection from a magnetar, \citealt{yu2013}). They become free after the ejecta reaches the transparent radius.
The corresponding X-ray light curve is expected to show a fast rise, followed by an evolution defined by magnetar wind dissipation.
There are three critical timescales to determine the shape of light curves: the magnetar spindown time $t_{\rm sd}$, the transparent time $t_{\tau}$, and the collapse time $t_{c}$. 
For $t_{\tau}<t_{\rm sd} \ll t_{c}$, the X-ray plateau phase (followed by a decay) will emerge during the rising phase. On the other hand, for 
$t_{\rm sd}<t_{\tau} \ll t_{c}$ a decaying light curve should immediately follow the rising phase. Depending on which component dominates the spindown, the temporal slope in the decay slope varies in the range of [-2,-1] \citep{sun2017}.

\subsection{Light curve in trapped zone}

The bolometric luminosity of merger-nova peaks in the ultraviolet/optical/infrared band. The merger-nova is usually dim in X-rays but can be bright under extreme conditions, e.g. an extremely large magnetic field, a very small initial spin period, or a low ejecta mass (\citealt{sun2017}, cf. \citealt{siegel2016a,siegel2016b}).
We calculate the merger-nova light curve following \cite{yu2013} and \cite{sun2017} (see those papers for detailed treatments).
The Lorentz factor of the ejecta evolves as
\begin{equation}
\frac{d\Gamma}{dt}=\frac{L_{\rm sd}+L_{\rm ra}-L_{e}-\Gamma{\cal D}(dE'_{\rm int}/dt')}{M_{\rm ej}c^2+E'_{\rm int}},
\end{equation}
where $L_{\rm sd}$, $L_{\rm ra}$, $L_e$ are the spindown luminosity from magnetar, radioactive heating luminosity and bolometric emission of merger-nova, respectively, $M_{\rm ej}$ is the ejecta mass, and $E'_{\rm int}$ is the internal energy of the ejecta in comoving frame. $ {\cal D}=1/[\Gamma(1-\beta\cos\theta)] $ is the Doppler factor, $ \beta=\sqrt{1-\Gamma^{-2}} $ is the dimensionless velocity, $\Gamma$ is the Lorentz factor, and $\theta$ is the viewing angle ($\theta = 0$ for an on-beam observer).

By tracking the comoving temperature of the ejecta $T'$, one can obtain the merger-nova luminosity at a given frequency $\nu$ as
\begin{equation}
(\nu L_{\nu})_{\rm bb}=\frac{8\pi^2 {\cal D}^2R^2}{h^3c^2}\frac{(h\nu /{\cal D})^4}{exp(h\nu/{\cal D}kT')-1},
\label{eq:Lmn}
\end{equation}
In general, the total  X-ray luminosity from the trapped zone can be calculated as
\begin{equation}
L_{\rm X,trapped}(t)= e^{-\tau} \frac{\eta B_p^2 R^6 \Omega^4(t)}{6c^3} + (\nu_{\rm X} L_{\rm \nu,X})_{\rm bb},
\label{eqs:X-trapped}
\end{equation}
where
\begin{equation}
\tau=\kappa (M_{\rm ej}/V')(R/\Gamma).
\end{equation} 
is the optical depth of the ejecta, $\kappa$ is the opacity of the ejecta for X-rays, and $V'$ is the comoving volume. The thermal component (the second term) is usually negligibly dim in X-rays. Therefore the luminosity in the trapped zone is dominated by the nonthermal component (the first term of Eq. \ref{eqs:X-trapped}).

\subsection{Pair production, ionization state, and opacity of the ejecta}
One complication in calculating the trapped-zone light curves is the optical depth for nonthermal X-rays. Pair production in the magnetar wind may increase the optical depth of X-ray photons, and the ionization state of the ejecta would significantly affect the opacity \citep{metzgerpiro2014}.

The effect of pair production can be evaluated by assuming the spectrum of nonthermal X-ray emission from magnetar wind dissipation, which is not well constrained from the observational data of SGRB X-ray plateau emission. Based on GRB phenomenology, we assume a broken-power-law spectrum similar to the Band function  of GRB prompt emission \citep{band1993}. This is because the X-ray plateau emission of SGRBs usually does not extend to gamma-rays, and if they do (detected by Swift BAT as the so-called ``extended emission'', the spectrum is quite soft. We assume an $E_{\rm peak} = 10$ $ \rm keV$, $\beta = -2.5$ and $\alpha = -\Gamma_p = -1.43$ for the band function and extend the spectrum to 0.511-10 $\rm MeV$ in the rest frame to estimate the pair production optical depth. This gives 
\begin{equation}
 \frac{L_{\gamma}}{L_{X}} \simeq \frac{\int_{0.511\rm MeV}^{10\rm MeV} EN(E)dE}{\int_{0.2\rm keV}^{10\rm keV}EN(E)dE} \simeq 0.18
\end{equation}
With the total energy $E_{\gamma} \simeq L_{\gamma}\Delta t$ and $\Delta t \simeq 100$ s, the optical depth for pair production is estimated as
\begin{eqnarray}
\tau_{\gamma \gamma} & = & n_{\gamma}\sigma_{T}R= \frac{\sigma_T L_{\gamma}\Delta tR}{\epsilon_{\gamma}(4/3)\pi R^3} 
 =  \frac{0.18 \sigma_T L_X}{\epsilon_\gamma (4/3) \pi c^2 \Delta t} \nonumber \\
& \simeq & 3\times 10^4 \left( \frac{L_X}{7\times 10^{46} \ {\rm erg \ s^{-1}}} \right)
\left(\frac{\epsilon_\gamma}{511 \ {\rm keV}} \right)^{-1} \left(\frac{\Delta t}{100 \ {\rm s}} \right)^{-1} \nonumber \\
& \gg & 1,
\end{eqnarray}
where we have adopted the following parameters: $R = c \Delta t$, $\Delta t \sim 100$ s,  $L_X$ is as adopted as the observed peak luminosity, and $\epsilon_\gamma = m_e c^2 = 511$ keV is the pair production threshold. Notice that this is a very conservative estimate. With the extreme magnetar parameters needed to fit the data of XT1, the ejecta would reach mild relativistic speeds. This would increase $R$ and decrease the photon energy in the comoving frame. Both effects lead to a decrease in $\tau_{\gamma \gamma}$, which drops below unity when the bulk Lorentz factor $\Gamma > 4.3$. 

Let us conservatively consider that pair production is important in the ejecta. One can compare the total number of produced pairs, $N_\pm$,  and the total number of electrons already in the ejecta, $N_e$. They can be estimated as\footnote{To calculate $N_\pm$, we have assumed that each 511 keV photon is converted to a lepton via $\gamma\gamma \rightarrow e^{+} e^{-}$. Photons with energy greater than 511 keV will interact with lower energy photons to produce pairs. Since the pair-producing $\gamma$-rays are in the $N(\epsilon)d\epsilon \propto \epsilon^{-2.5} d \epsilon$ regime, the pairs are predominantly produced at the energy $\epsilon_\gamma$=511 keV.}
\begin{equation}
 N_{\pm} \simeq \frac{L_{\gamma}\Delta t}{\epsilon_{\gamma}} \simeq 1.5 \times 10^{54} \left( \frac{L_X}{7\times 10^{46} \ {\rm erg \ s^{-1}}} \right)
 \left(\frac{\Delta t}{100 \ {\rm s}} \right), 
\end{equation}
and
\begin{equation}
 N_{e} \simeq Y_e \frac{M_{\rm ej}}{m_p} = 4.8 \times 10^{53}\left(  \frac{M_{\rm ej}}{10^{-3}M_{\odot}}\right) \left(\frac{Y_e}{0.4} \right),
\end{equation}
respectively, where $Y_e$ is the electron fraction. Since the pair multiplicity ${\cal M}_e \equiv N_\pm / N_e \simeq 3$, one can see that pair production would moderately increase the opacity of X-rays in the ejecta.

Next, one can check the ionization state of the ejecta. With a millisecond magnetar as the central engine, the ejecta is continuously illuminated by the X-ray flux from the magnetar. The ionization energy for the innermost electronic state for Fe is $\sim 9.3$ keV. For $E_p \sim 10$ keV, the X-ray number flux at 10 keV is
\begin{equation}
 \dot N_X = L_X/(10 \ {\rm keV}) = 4.4 \times 10^{54} \ {\rm s^{-1}} \ \left( \frac{L_X}{7\times 10^{46} \ {\rm erg \ s^{-1}}} \right).
 \label{eq:ionization}
\end{equation}
One can see that the ejecta can be fully ionized by this strong X-ray flux within 0.1 s for $M_{\rm ej} = 10^{-3} M_\odot$, and in around 1 s even for  $M_{\rm ej} = 10^{-2} M_\odot$. 
In the following, we assume that the ejecta is fully ionized.\footnote{In order to access whether the ejecta can be fully ionized, one also needs to consider the recombination time scale of the ejecta. 
The time scale of recombination depends on density and recombination coefficients, which further depends on the ionization state of Fe and temperature \citep{woods1981}.  The ionization rate also depends on the ionization state with the ionization energy ranging from 7.9 $\rm eV$(Fe I ) to 9.3 $\rm keV$ (Fe XXVI). Detailed calculations using numerical tools such as CLOUDY \citep{ferland2017} are needed to further justify the full ionization hypothesis.}

The opacity of X-rays is dominated by electron Thomson scattering for a fully ionized ejecta. The X-ray opacity can be estimated as
\begin{equation}
 \kappa = \frac{\sigma_T n_e} {\rho} = \frac{\sigma_T \xi_e}{m_p} \simeq (0.6 \ {\rm  cm^{2}g^{-1}}) \frac{\xi_e}{1.6},
\end{equation}
where $\xi_e = (1+{\cal M}_e) Y_e \sim 1.6$ for our nominal parameters. In our following modeling, $\kappa = 1 \ {\rm  cm^{2}g^{-1}}$ is approximately adopted. 

\section{The light-curve fit}

Comparing the observed light curves with the model above, one can infer that XT1 likely originates from the trapped zone while XT2 originates from the free zone. We perform the light-curve fitting with the least square method. The free parameters include $B_p$, $P_i$ and $\eta$ (for both events) and  additional parameters of $M_{\rm ej}$ and the zero time point $T_0$ for XT1. In \cite{bauer2017},  $T_0$ is poorly constrained and was arbitrarily set as 10 s prior to the arrival of the first photon and $\sim 150$ s before the peak \citep{bauer2017}. This gap could be larger if the early emission is obscured. It is better constrained for XT2 as no photon was detected 10 s prior to the peak luminosity \citep{xue2019}. 

\begin{figure}[hbtp]
\centering
\includegraphics[width=0.5\textwidth]{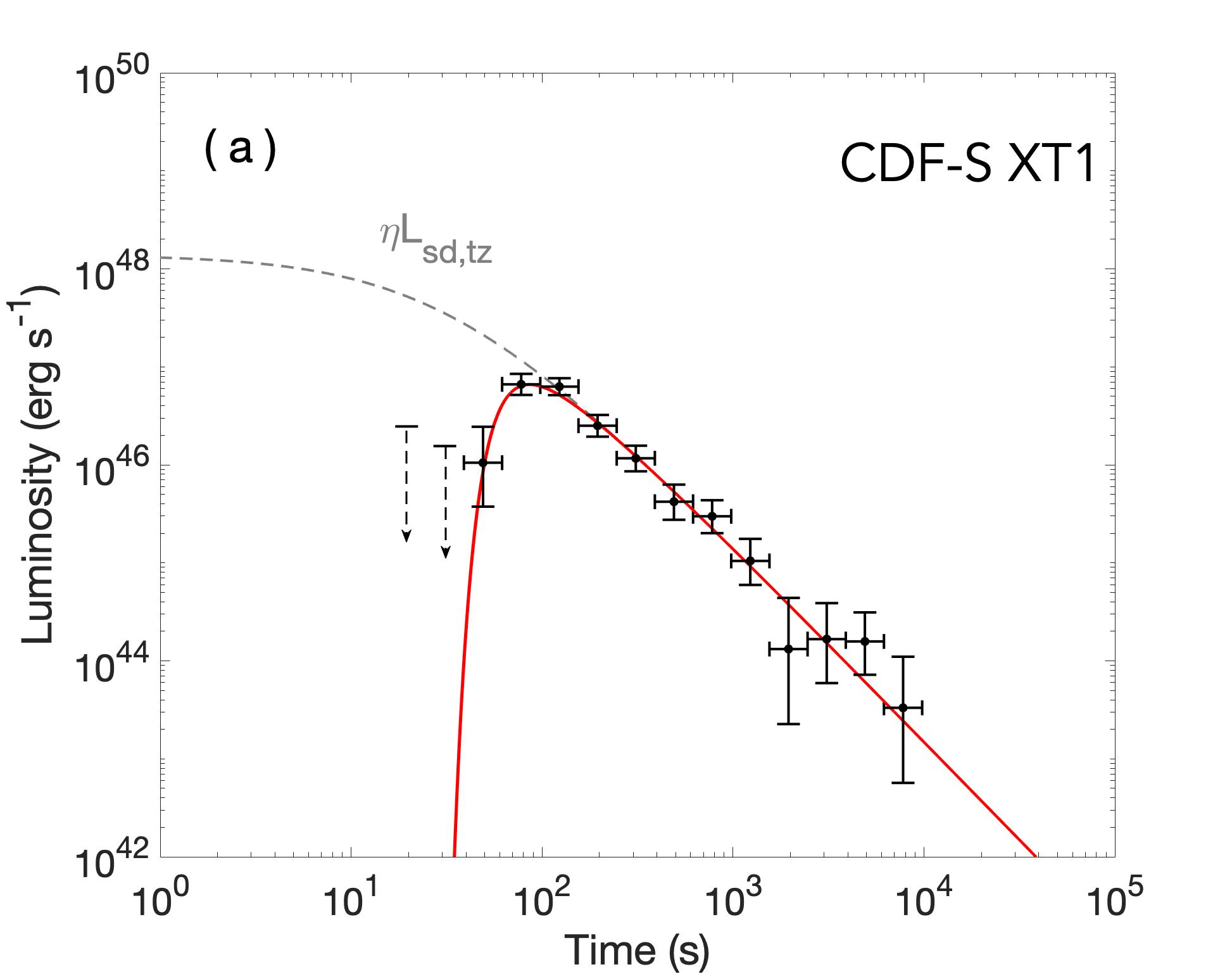}
\includegraphics[width=0.5\textwidth]{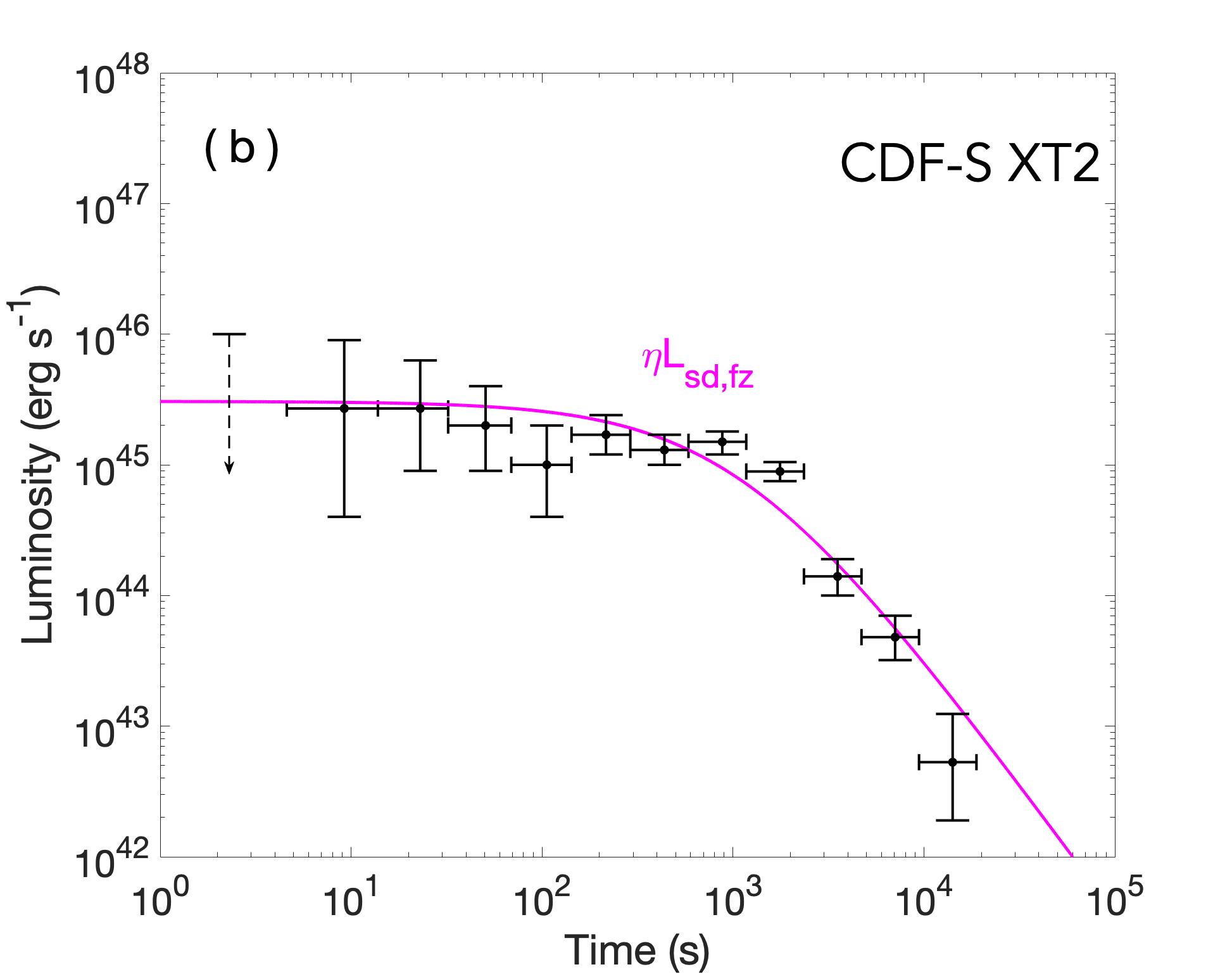} 

\caption{Light-curve fitting of luminosities in the source rest frame for the two transients, CDF-S XT1 (a) and XT2 (b). The black dots with errors are the binned data with time dilation corrected. (a) For XT1, the data are taken from \cite{bauer2017}.  The red curve is an example good fit with the trapped zone model, with the unabsorbed trapped zone luminosity ($\eta L_{\rm sd,tz}$) marked in gray. 
(b) For XT2, the data are taken from \cite{xue2019}, and an example free zone fit is shown as the magenta curve. }
\label{Fig:fit_lc}
\end{figure}

The light-curve fitting is presented in Figure \ref{Fig:fit_lc}, with the results summarized in Table \ref{Tab:fit_lc}. The  rest-frame 0.3-10 $\rm keV$ luminosities are presented for both transients, with XT1 derived from $k$-correction assuming a spectral slope of 1.43 and XT2 directly from \cite{xue2019}. For XT1 (upper panel of Figure \ref{Fig:fit_lc}), the light-curve data points can be generally reproduced with the trapped zone geometry (red line). 
An example good fit (with $\chi^2$/dof close to unity) gives $B_p=10^{16}\,G$,  $P_i=1.2$ $\rm ms$, $M_{\rm ej}=0.001\, M_{\odot}$, $\eta = 0.001$ and $T_0 = -140$ s.
The fast-rising phase corresponds to the emergence of the X-ray emission produced from magnetar wind dissipation as the optical depth drops with time, with the peak at the epoch $\tau \sim 1$. 
The decline phase of the XT1 light curve can be well fitted by the spindown luminosity after the spindown time scale up to 10 ks. The ejecta mass is fitted to be $10^{-3}$  $M_\odot$.

For XT2 (the lower panel of Figure \ref{Fig:fit_lc}), the light curve can be well fitted by the free-zone model (magenta line) with $B_p=10^{15.8}\,G$, $P_i=4.4$ $\rm ms$, $\eta = 0.001$, which is consistent with the modeling of this event \citep{xue2019,xiao2019,lv2019}. The $\rm \chi ^2/dof$ is slightly over 2 but still leads to a relatively small p-value ($p\sim 0.012$ at the significance of 0.05). The low efficiency of $\sim 10^{-3}$ for both XT1 and XT2 is theoretically expected within the slow magnetic wind dissipation model with a saturation Lorentz factor $\Gamma_{\rm sat} = 10^3$-$10^4$
\citep{ xiao2019}.  

\begin{figure}[hbtp]
\centering
\includegraphics[width=0.45\textwidth]{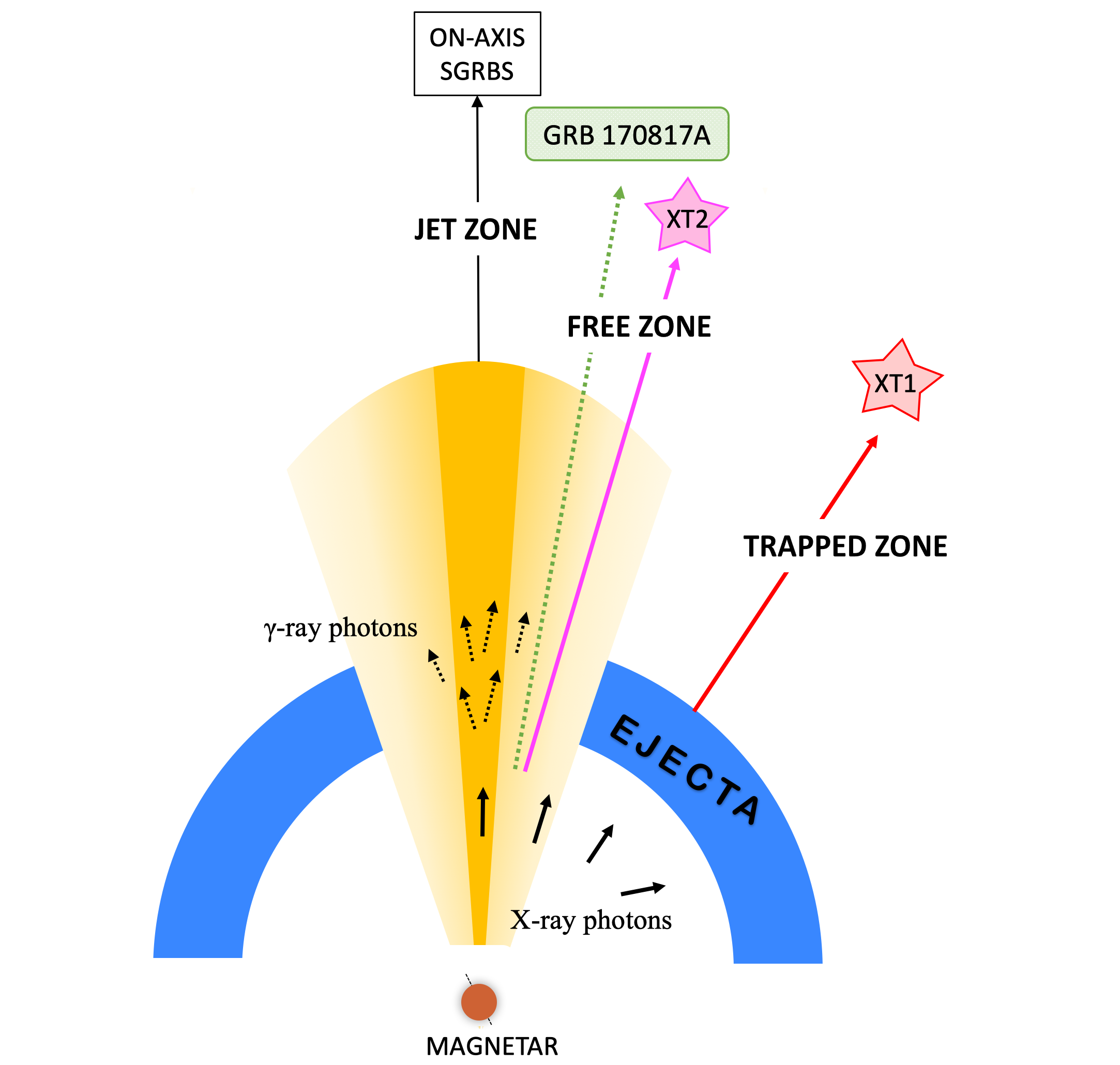} 
\caption{Schematic diagram of the geometry highlighting three emission (jet, free and trapped) zones and the possible viewing geometries of XT1 and XT2. A structured jet is shown with gradients in yellow.  
XT2 (magenta) is likely observed from the free zone with a viewing angle similar to GRB 170817A (green) . XT1 (red) is more likely from the trapped zone.}
\label{Fig:fit_sch}
\end{figure}

It is interesting to note that the fitting results of XT1 and XT2 suggest comparable magnetar parameters (e.g., magnetic field, initial spin period and transfer efficiency) for the two transients, implying a unified origin for the two transients with different viewing geometries (summarized in Figure \ref{Fig:fit_sch}). 

Similar to \cite{xue2019}, who placed an upper limit on the gamma-ray luminosity of a putative SGRB $L_{8-100 \rm keV} = 3.5 \times 10^{49}$ $\rm erg\,s^{-1}$ for XT2, 
we also conducted a search for a possible gamma-ray signal associated with XT1 using the {\em Fermi}-GBM data. We searched the signal in a time interval from $-10^4$ to $2\times 10^4$ s around $T_0$ (from \cite{bauer2017}) and found no significant transient above 10 keV.
The 1-$\sigma$ flux upper limit in 8-100 keV range between -50 s and 50 s is $4.73^{+8.19}_{-4.67}\times 10^{-9}$ $\rm erg\, cm^{-2}\,s^{-1}$ assuming a power-law spectrum. The 1-$\sigma$ flux upper limit in the same energy range between 200 s and 500 s is $1.2^{+0.35}_{-0.35}\times 10^{-8}$ $\rm erg\, cm^{-2}\,s^{-1}$. Both flux upper limits lead to an upper limit luminosity of $\sim 10^{50}$ $\rm erg\,s^{-1}$, which is higher than the luminosity of GRB 170817A but lower than those of the majority of on-axis SGRBs.

\renewcommand\arraystretch{2}

\begin{table*}[t]
\caption{Summary of properties of XT1 and XT2  and fitting results.}
\centering
\begin{tabular}{c|c|c|c|c|c|c|c|c|c|c}
\hline 
\hline 
\rule[-1ex]{0pt}{2.5ex}  & Viewing Direction   &Redshift  & $\rm L_{p}/(\rm erg\,s^{-1})$& $\Gamma_{p}$ &$B_p/(\rm G)$ & $P_i/(\rm ms)$  & $\eta$ & $M_{\rm ej}/(M_{\odot})$& $T_0/(s)$& $\rm \chi^2/dof$\\ 
\hline 
\rule[-1ex]{0pt}{2.5ex} XT1 &Trapped Zone & 2.23($z_{\rm ph} $)  &$6.8\times 10^{46}$ &1.43  &  $10^{16}$ & 1.2 &0.001 & 0.001&-140 &$4.15/7$\\ 
\hline 
\rule[-1ex]{0pt}{2.5ex} XT2&Free Zone  &0.738 & $2.7\times 10^{45}$&1.93 & $10^{15.8}$ & 4.4 & 0.001&-&-&$19.48/8$\\
\hline 
\end{tabular} 
\label{Tab:fit_lc}
\end{table*}

The ejecta mass of XT1 in our example fit, $M_{\rm ej} = 10^{-3} \ M_{\odot}$ falls into the range of the ejecta mass from numerical simulations of binary neutron star mergers \citep[e.g.][]{hotokezaka2013}. On the other hand, recent simulations (mostly prompted by the observation of the bright kilonova associated with GW170817) show that the ejecta mass could be as high as $10^{-2} \ M_{\odot}$, especially when the strong neutrino-driven disk wind is taken into account \citep{just2015,siegel2017,fernandez2019b}. Our example $M_{\rm ej}$ seems lower than this. We notice that the modeling presented here, unlike the kilonova modeling, does not constrain the total $M_{\rm ej}$, but only constrains the ``isotropic equivalent'' value of $M_{\rm ej}$ along the line of sight. This is because the non-thermal X-ray emitter (the magnetar wind) is moving relativistically. Once the line-of-sight ejecta becomes transparent, the X-ray flux would rise significantly. Since the ejecta mass is not distributed isotropically, it is possible to derive an effective $M_{\rm ej}$ smaller than the true ejecta mass if the line of sight is not too deep into the trapped zone. 

\section{Comparison with SGRBs}

Since SGRBs are believed to originate from BNS mergers, as manifested in the GW170817/GRB 170817A association, we perform a comparison of the two X-ray transients (XT1 and XT2) with the previously observed SGRBs.

\subsection{Magnetic fields and initial spin periods} 

A fraction (1/3 to 1/2) of SGRBs detected by {\em Swift} are followed by extended emission or an X-ray plateau, which can be interpreted as the emission of the dissipating wind of a post-merger magnetar \citep{rowlinson2013,lv2015}. Since the young magnetar wind is essentially isotropic,
it is expected that SGRB-less X-ray transients similar to XT1 and XT2 exist and should share similar properties as the X-ray plateaus \citep{zhang2013}. 

\begin{figure}[hbtp]
\centering
\includegraphics[width=0.5\textwidth]{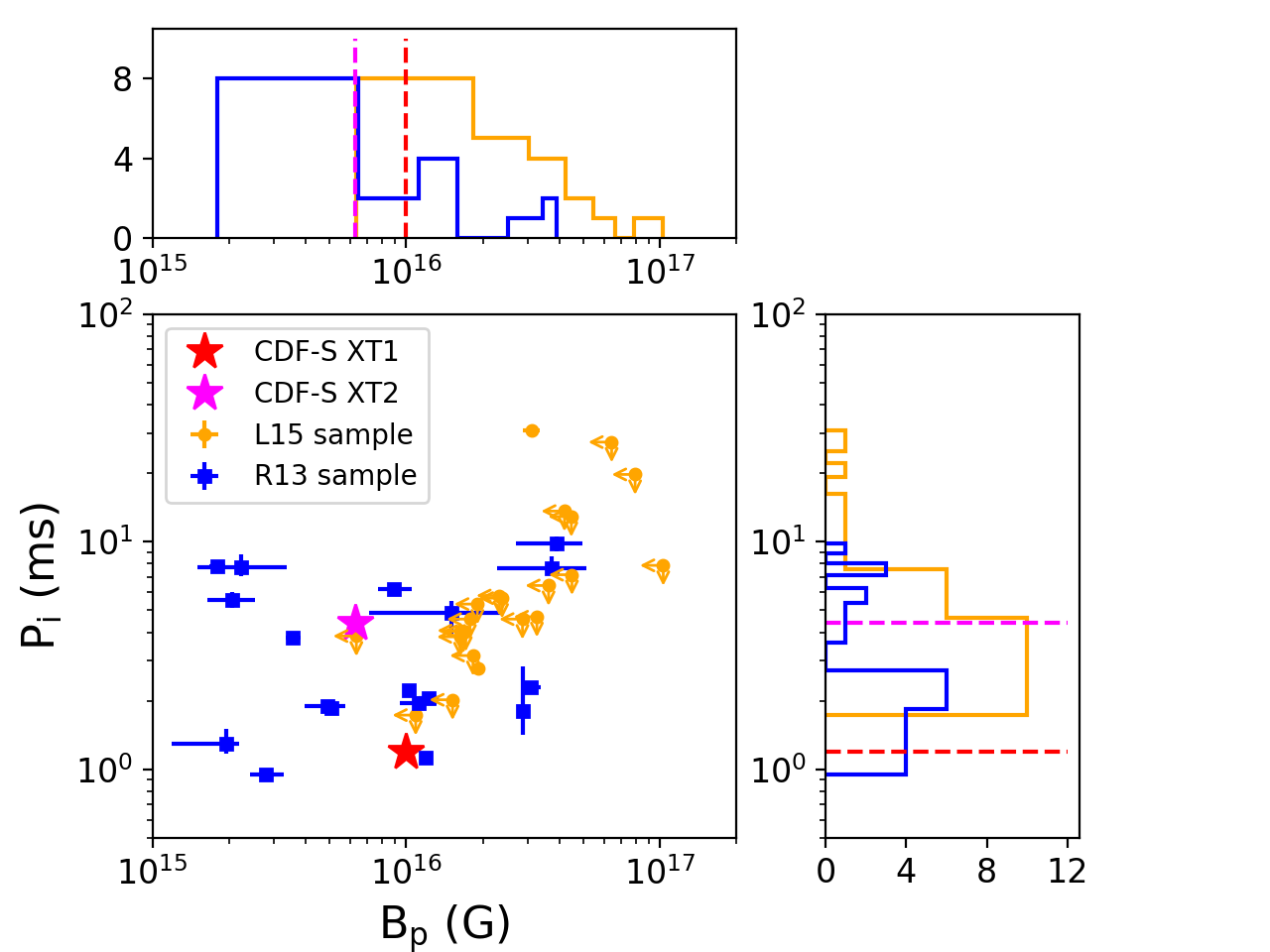} 
\caption{Comparison of the derived magnetar parameters ($B_p$ and $P_i$) of XT1/XT2 with SGRB plateaus.
Blue squares mark the magnetar sample from R13. The orange dots mark the magnetar sample from L15, among which those with arrows represent the supramassive population which collapse to black holes so that the data only give the upper limits of $B_p$ and $P_i$. 
The upper and right panels are the histograms for $B_p$ and $P_i$, respectively, with the red and magenta dashed lines marking the transients XT1 and XT2, respectively. The upper limits from L15 are directly taken as face values for the orange histograms.} 
\label{Fig:fit_bp}
\end{figure}

We first compare the derived magnetar parameters ($B_p$ and $P_i$) of the SGRB X-ray plateau population and those of XT1 and XT2. For the SGRBs, we adopt the sample of L15 and R13 and their derived magnetar parameters. Since the L15 sample includes many plateaus followed by a steep decay segment (the so-called internal plateaus and best interpreted as the collapse of an SMNS at the end of the plateau), the derived $B_p$ and $P_i$ are only upper limits for these events. 
Only two candidates in the sample are SNSs with the true fitted values of $B_p$ and $P_i$. 
As shown in Figure \ref{Fig:fit_bp}, the SGRB magnetar sample typically has a $B_p$ distribution between $10^{15}\,G$ and $10^{17}\,G$ and a $P_i$ distribution between 1 ms and 10 ms. The values for XT1 and XT2 both fall into these ranges, suggesting a similar origin.

\subsection{Photon indices}

\begin{figure}[hbtp]
\centering
\includegraphics[width=0.4\textwidth]{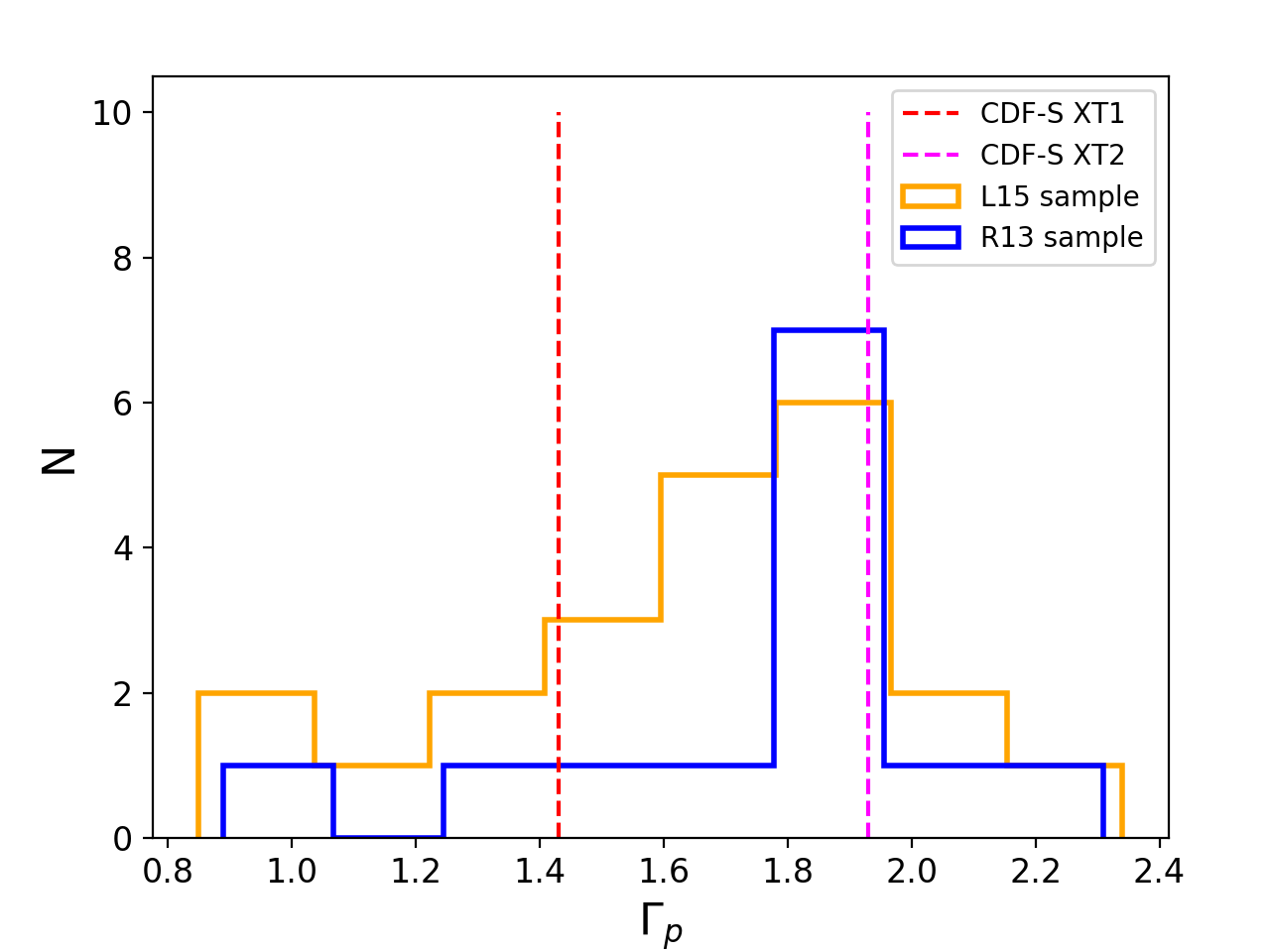} 
\caption{Comparison of the average photon indices of XT1/XT2 with SGRB plateaus. Red and magenta dashed lines mark the XT1 \citep{bauer2017} and XT2 \citep{xue2019}, respectively. The blue and orange histograms are for the samples of R13 and L15, respectively.}
\label{Fig:fit_gm}
\end{figure}

In Figure \ref{Fig:fit_gm}, we compare the photon indices of XT1 and XT2 with those of the SGRB X-ray plateaus. Both the R13 sample ($\Gamma_{X,2}$ for GRBs with two or more breaks and $\Gamma_X$ in the segment with a shallower $\alpha$ for GRBs with one break) and the L15 sample (BAT 15-150 $\rm keV$ photon index $\Gamma_{\gamma}$ for the extended emission sample and XRT-band photon index for the X-ray plateau sample) are adopted. One can see that the photon index of XT2 is typical for SGRBs. Even though that of XT1 is not typical, it nonetheless falls within the SGRB photon index distribution.

\subsection{Host-galaxy properties}
\begin{figure}[hbtp]
\centering
\includegraphics[width=0.45\textwidth]{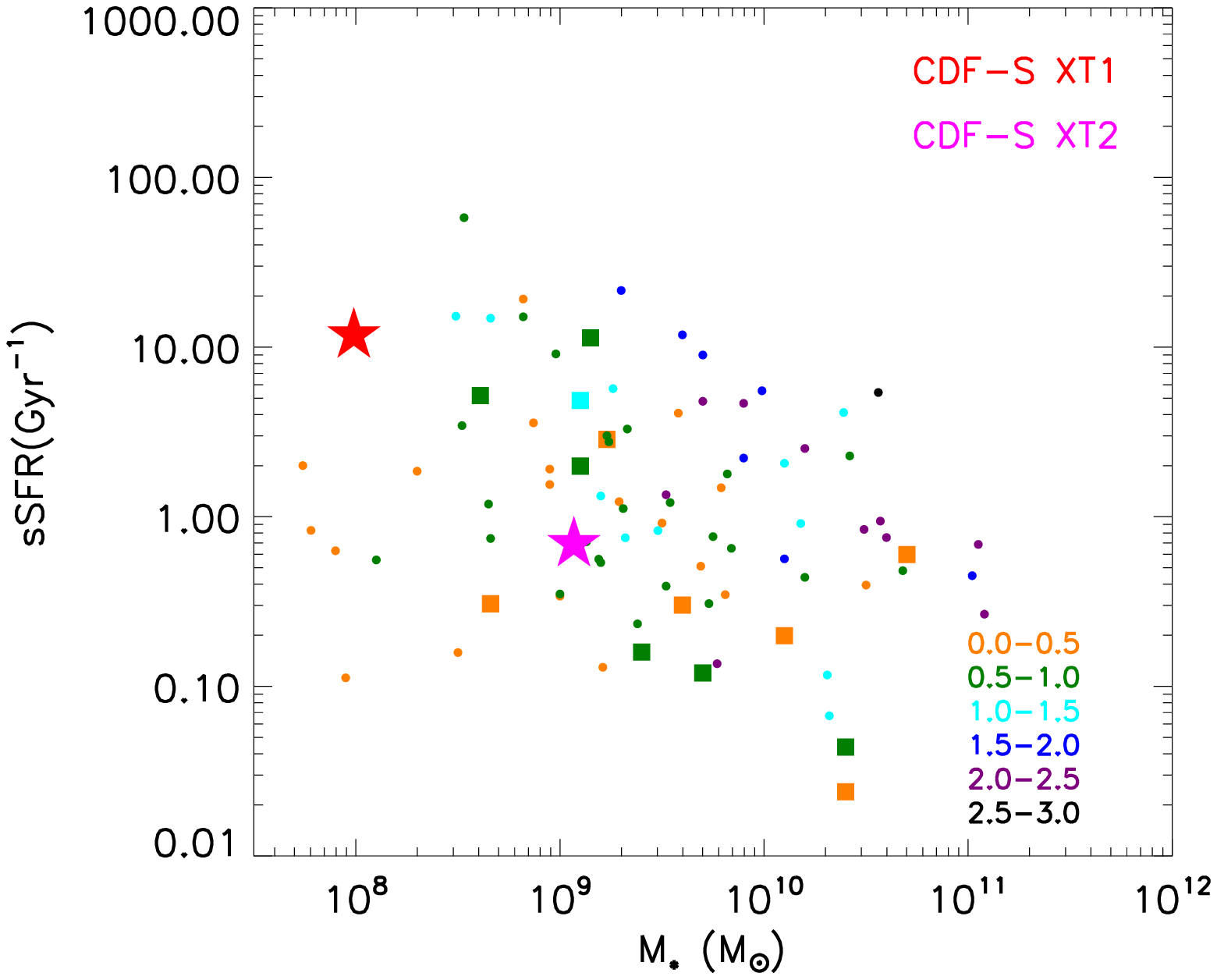} 
\includegraphics[width=0.45\textwidth]{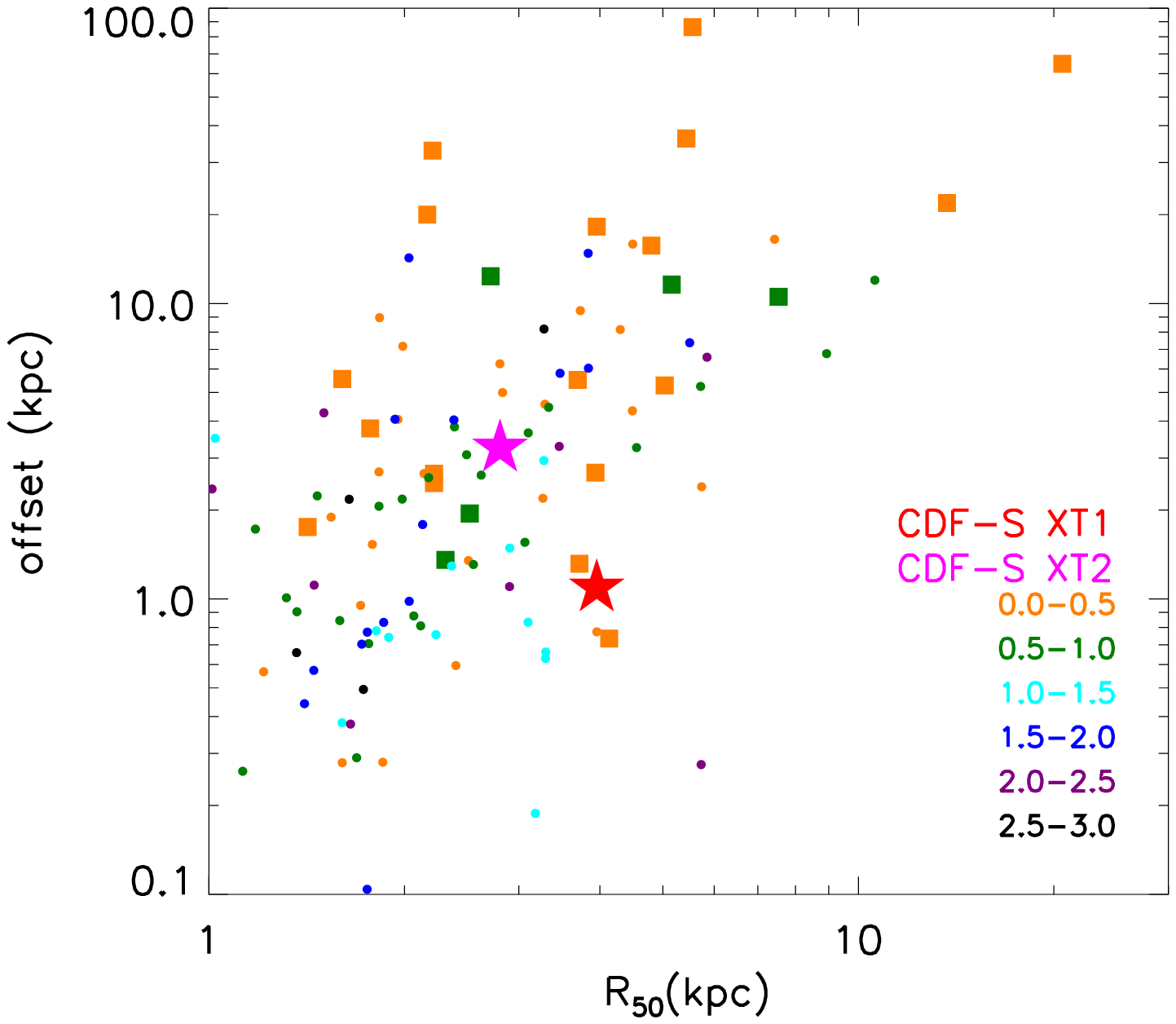} 
\includegraphics[width=0.45\textwidth]{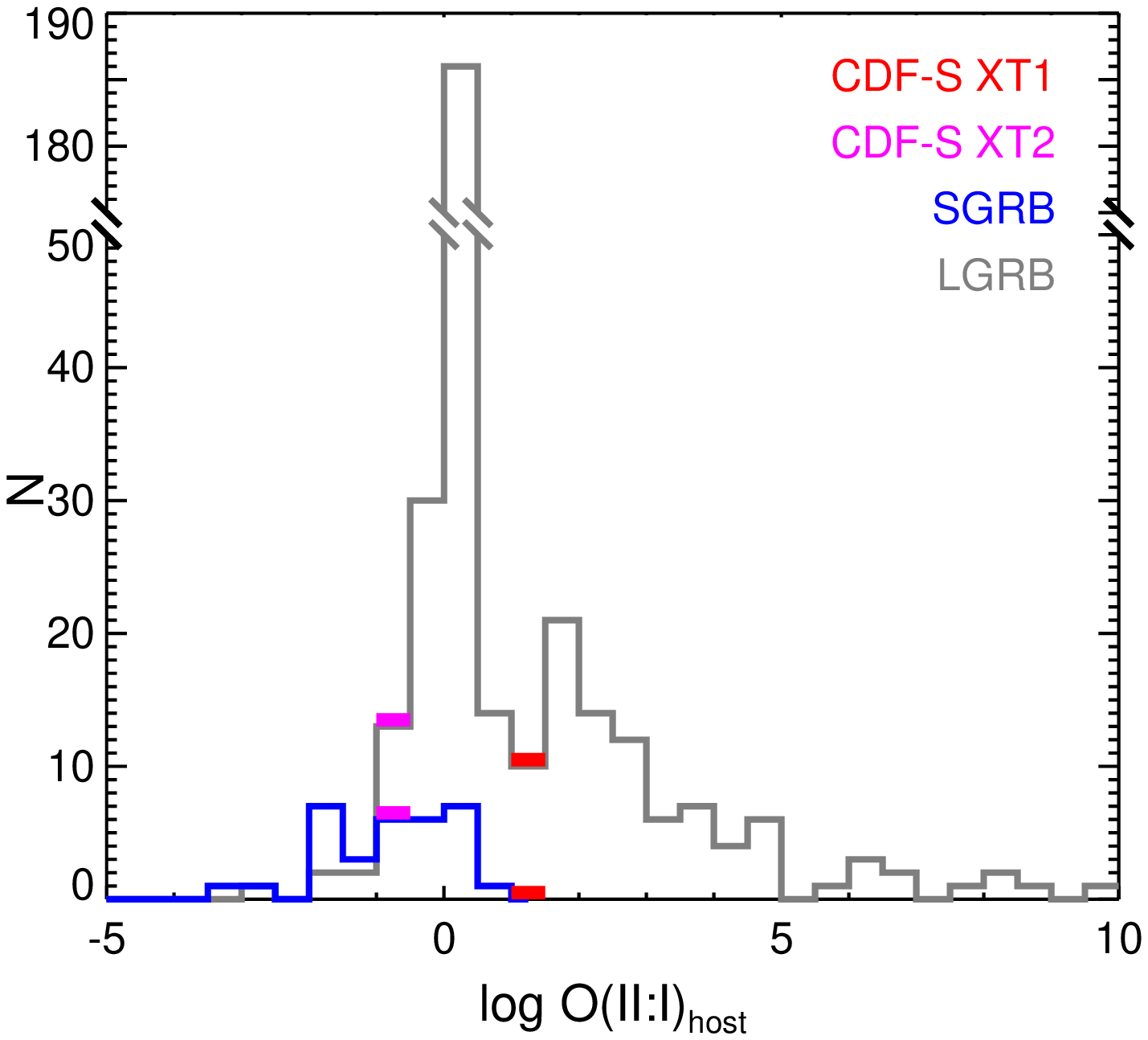} 
\caption{Comparison of the host-galaxy properties of XT1/XT2 with SGRBs (squares) and LGRBs (dots). Different colors mark different redshift ranges. The red/magenta stars represent the host galaxies of XT1/XT2. The upper panel shows the distributions in the sSFR vs. $M_*$ domain. The middle panel shows the offset vs. $R_{50}$, both in units of $\rm kpc$. The low panel shows the histograms of $\rm O(II:I)_{host}$. }
\label{Fig:fit_galaxy}
\end{figure}

We compare the host-galaxy properties of XT1 and XT2 with those of redshift-binned SGRBs and long GRBs (LGRBs) in Figure \ref{Fig:fit_galaxy}. The host galaxy data of GRBs are adopted from \cite{liye2016} and references therein \citep[e.g.,][]{berger2009,fong2013,blanchard2016}. They are presented in both the specific star formation rate (sSFR) - stellar mass ($M_*$) plane and the offset - half-light radius $R_{50}$ plane. We only include SGRBs with SED-estimated stellar masses and emission-line estimated SFRs in the $M_*$ vs. sSFR scatter plot. The host properties of XT2 resemble those of SGRBs at similar redshift ($z \sim$ 0.7) in all four quantities.
For XT1, the host-galaxy mass appears to be smaller than any among the known SGRBs, while the SFR is comparable to them. Since the redshifts of all known SGRBs are (much) smaller than photometric redshift $z_{ph}=2.23$ of XT1 \citep{bauer2017}, and since galaxies are typically smaller at high redshifts \citep{bouwens2004, mosleh2012}, the XT1 host may not be regarded as abnormally small compared with the SGRB sample. 
The galaxy size $R_{50}$ of XT1 is estimated from the Kron radius $r_{\rm kron}$ as $R_{50}=r_{\rm kron}/1.19$ by assuming the Sersic index $n=1$, which is generally consistent with late-type galaxies. The $R_{50}$ and the offset of XT1, in units of kpc, are estimated with $z_{ph}=2.2$3 \citep{bauer2017}. The offset of XT1 belongs to the lower end of the offset distribution of SGRBs.
The lower panel of Figure \ref{Fig:fit_galaxy} describes $\rm O(II:I)_{host}$, representing the ``odds'' or probabilities that the sources belong to the LGRB (massive-star core collapse type, or Type II) vs. SGRB (compact-star merger type, or Type I) populations based on the statistical properties of the host-galaxy data of the two types \citep{liye2016}. As pointed out by \cite{xue2019}, the $\rm O(II:I)_{host}$ of XT2 falls right on the peak of the distribution of SGRBs. We similarly calculate $\rm O(II:I)_{host}$ for XT1 and find that it can in principle belong to the SGRB population (high end of distribution) even though it is also consistent with the LGRB population.

\section{Event rate density}

Based on GW170817 detection, the local event rate density of NS-NS mergers is estimated as $1.5^{+3.2}_{-1.2}\times 10^3$ $\rm Gpc^{-3}\,yr^{-1}$ \citep{abbott2017a, zhang2018}. It would be interesting to check how the event rate densities of XT1 and XT2 compares with this rate. 

The event rate density in general evolves with redshift as $\rho(z) = \rho_0f (z)$. \cite{sun2015} derived the redshift evolution function $f(z)$ of NS-NS mergers considering the merger delay distribution with respect to the star formation history. The local event rate density $\rho_0$ can be estimated via $N=\rho_0 V_{\rm max} \Omega T/(4\pi)$, where $V_{\rm max}$ represents the maximum volume a transient like XT1 or XT2 can be detected by CDF-S,
and $\Omega$ and $T$ represent the field of view (FOV) and on-sky exposure time, respectively. Even though both transients were discovered from CDF-S, different search strategies were performed by the two discovery teams. Different criteria lead to different limits of sky coverage and exposure time. 

For the peak flux of XT1, Chandra can detect similar events even if they are fainter by a factor of 10. Therefore, given the peak luminosity of $L_{\rm 2-10\,keV}=$ $6.8 \times 10^{46}$ $\rm erg\,s^{-1}$, Chandra can detect similar events to a much higher redshift ($z_{\rm max}=3.2$), with a maximum redshift-corrected comoving volume \footnote{ Here we take the SGRB redshift distribution considering a Gaussian merger model, i.e., Equation (20) in \citealt{sun2015}.} $\sim 3000$ $\rm Gpc^3$. 
The total on-sky exposure time given by \cite{bauer2017} is the combination of 46.6 Ms for four of the six ACIS-I detectors, with a FOV of $\rm 289\, arcmin^2$, 62.1 Ms for three ACIS-S detectors with a FOV of $\rm 217\, arcmin^2$, and 3.7 Ms of central $\rm 100\, arcmin^2$ for HRC. This gives $\rho_{0,\rm XT1} =54^{+124}_{-45}$ $ \rm Gpc^{-3}\,yr^{-1}$ (with 1$\sigma$ errors hearafter). This is much smaller than $\rho_{0,\rm BNS}$. 

For the peak flux of XT2, Chandra can detect similar events up to $\sim z_{\rm max}= 1.9$, with a maximum redshift-corrected comoving volume $\sim 2200$ $\rm Gpc^3$. The total on-sky exposure time given by \cite{xue2019} is 7 Ms, with a FOV of 0.05 square degrees. This gives\footnote{It is consistent with the event rate density derived by \cite{xue2019}, i.e. $\rho_{0,\rm XT2} =1.8^{+4.1}_{-1.6}\times 10^3$ $ \rm Gpc^{-3}\,yr^{-1}$.} $\rho_{0,\rm XT2} =1.4^{+3.3}_{-1.2}\times 10^3$ $ \rm Gpc^{-3}\,yr^{-1}$. The XT1 was also discovered during the search for XT2. The corresponding event rate density of XT1 following this strategy is $\rho_{0,\rm XT1} =1.1^{+2.5}_{-0.9}\times 10^3$ $ \rm Gpc^{-3}\,yr^{-1}$. Both the event rate densities of the XT1 and XT2 are consistent with $\rho_{0,\rm BNS}$. 

In order to make transients XT1 and XT2, a BNS merger needs to leave behind either an SMNS or an SNS. Based on various constraints, this fraction is generally above 60\% \citep{gao2016,margalit19}. Our example fits suggest that the last observed data point for both cases is after the characteristic spindown time scale. Since there is no evidence of magnetar collapse from the light curve (which would appear as a much steeper decay), there is a high probability that the remnant for both cases is an SNS. There is a disagreement on the fraction of SNS remnants for BNS mergers. \cite{margalit19} suggested that it is less than 3\% based on the assumption that the merger remnant of GW170817 is a BH followed by a brief HMNS phase and that the maximum mass of a non-spinning neutron star ($M_{\rm TOV}$) is 2.17 $M_\odot$ \citep{margalitmetzger2017,ruiz2018,rezzolla2018}. If this is the case, our interpretation of XT1 and XT2 as BNS merger remnants would be challenged. On the other hand, this low $M_{\rm TOV}$ is inconsistent with the short GRB X-ray plateau data, which suggest $M_{\rm TOV} > 2.3 M_\odot$ and SNS fraction $\sim 30\%$ \citep{gao2016}. The merger remnant of GW170817 is not well settled (cf. \citealt{ai2018,yu2018,li2018,piro2019}). Even if the merger product is a BH, \cite{shibata19} recently showed that $M_{\rm TOV}$ upper limit should be $2.3 M_\odot$ rather than 2.17 $M_\odot$. This is close to the value from the short GRB X-ray plateau constraint. If so, the derived event rate densities from XT1 and XT2 are still broadly consistent with BNS mergers with an SNS remnant.

The fact that the event rate density of XT1 is lower than that of XT2 may seem unexpected, since the trapped zone solid angle is typically much larger than that of the free zone \citep{sun2017}. On the other hand, the majority of the trapped zone X-ray transients should have much lower luminosities than XT1. In order to be still detectable in the trapped zone, the magnetar parameters should be even more extreme so that the plateau luminosity is higher. Indeed XT1 has a stronger $B_p$ and a shorter $P_i$ than XT2, suggesting more extreme conditions. The lower event rate density of XT1 is therefore understandable.

\section{Conclusions}

In this paper, we have proposed a unified magnetar model of BNS merger to explain both X-ray transients discovered from the Chandra Deep Field-South Survey, i.e. CDF-S XT1 and CDF-S XT2. The model can explain well the observed light curves of the two events, with XT1 from the trapped zone (fitting parameters $B_p=10^{16}\,G$, $P_i=1.2\, \rm ms$, $\eta = 0.001$, $M_{\rm ej}=0.001\, M_{\odot}$, $T_0 =$ -140 s) and XT2 from the free zone (fitting parameters $B_p=10^{15.8}\,G$, $P_i=4.4\, \rm ms$, $\eta = 0.001$).
This suggestion is supported by the consistency of the properties of these two events with those of the SGRB X-ray plateaus in terms of light curves and spectra as well as the host-galaxy properties. The estimated event rate densities of these transients are consistent with that of BNS mergers \citep{abbott2017a,zhang2018}, suggesting that most of the BNS mergers may have left behind long-lived massive neutron stars. The X-ray opacity which was assumed to be 1 $\rm cm^2\,g^{-1}$ in this work needs to be proven by future simulations.

It is predicted that future multimessenger observations of BNS merger events may catch more such X-ray transients associated with BNS gravitational wave events. Whereas SGRBs are beamed, the magnetar-powered X-ray transients have much wider solid angles, so that most BNS mergers may be associated with a SGRB-less X-ray transient \citep{zhang2013,sun2017}. Such transients are not easy to detect with the current wide-field GRB detectors but could be ideal targets for 
future wide field X-ray telescopes such as the Einstein Probe  \citep{yuan2016}. The detections of these transients in the future will play a key role in identifying the BNS merger remnant, constraining the NS equation of state, and probing the energy power of kilonova/merger-nova. 

Finally, even though we proposed a unified model for CDF-XT1 and XT2, it is still possible that the two have distinct origins. Furthermore, it may be possible to interpret both events with scenarios other than BNS origins. For example, \cite{fernandez2019a} argued that XT2 can also be interpreted as the merger product of a white dwarf and a neutron star or black hole. \cite{peng2019} suggested that tidal disruption of white dwarfs by an intermediate black hole may be possible to interpret both XT1 and XT2. Our proposed model can be eventually confirmed via joint GW / X-ray detections in the future.

\acknowledgments
We thank the referee for important comments. H. Sun thanks helpful discussions on the magnetar sample with Houjun L\"u and on lightcurve fitting with He Gao. H.S. and W.M.Y. acknowledge support by the Strategic Pioneer Program on Space Science, Chinese Academy of Sciences, Grant No.XDA15052100 and the Strategic Priority Research Program of the Chinese Academy of Sciences Grant No. XDB23040100. Y.L. is supported by the KIAA-CAS Fellowship, which is jointly supported by Peking University and Chinese Academy of Sciences. This work is also partially supported by the China Post-doctoral Science Foundation (No. 2018M631242). B.-B.Z. acknowledges support from National Thousand Young Talents program of China, the National Key Research and Development Program of China (2018YFA0404204), and NSFC-11833003. F.E.B. acknowledges support from CONICYT-Chile Basal AFB-170002 and the Ministry of Economy, Development, and Tourism's Millennium Science Initiative through grant IC120009, awarded to The Millennium Institute of Astrophysics, MAS. Y.Q.X. acknowledges support from NSFC grants (11890693 and 11421303).

{}


\begin{thebibliography}{}
\bibitem[Abbott et al.(2017a)]{abbott2017a} Abbott, B.~P., Abbott, R., Abbott, T.~D., et al.\ 2017a, Physical Review Letters, 119, 161101 
\bibitem[Abbott et al.(2017b)]{abbott2017b} Abbott, B.~P., Abbott, R., Abbott, T.~D., et al.\ 2017b, \apjl, 848, L12 
\bibitem[Ai et al.(2018)]{ai2018} Ai, S., Gao, H., Dai, Z.-G., et al.\ 2018, \apj, 860, 57
\bibitem[Band et al.(1993)]{band1993} Band, D., Matteson, J., Ford, L., et al.\ 1993, \apj, 413, 281 

\bibitem[Bauer et al.(2017)]{bauer2017} Bauer, F.~E., Treister, E., Schawinski, K., et al.\ 2017, \mnras, 467, 4841 
\bibitem[Berger(2009)]{berger2009} Berger, E.\ 2009, \apj, 690, 231 
\bibitem[Blanchard et al.(2016)]{blanchard2016} Blanchard, P.~K., Berger, E., \& Fong, W.-f.\ 2016, \apj, 817, 144 
\bibitem[Bouwens et al.(2004)]{bouwens2004} Bouwens, R.~J., Illingworth, G.~D., Blakeslee, J.~P., et al.\ 2004, \apjl, 611, L1
\bibitem[Cromartie et al.(2019)]{cromatie2019} Cromartie, H.~T., Fonseca, E., Ransom, S.~M., et al.\ 2019, Nature Astronomy, 439
\bibitem[Dai et al.(2006)]{dai2006} Dai, Z.~G., Wang, X.~Y., Wu, X.~F., \& Zhang, B.\ 2006, Science, 311, 1127 
\bibitem[Ferland et al.(2017)]{ferland2017} Ferland, G.~J., Chatzikos, M., Guzm{\'a}n, F., et al.\ 2017, \rmxaa, 53, 385

\bibitem[Fern{\'a}ndez et al.(2019a)]{fernandez2019a} Fern{\'a}ndez, R., Margalit, B., \& Metzger, B.~D.\ 2019a, \mnras, 488, 259
\bibitem[Fern{\'a}ndez et al.(2019b)]{fernandez2019b} Fern{\'a}ndez, R., Tchekhovskoy, A., Quataert, E., et al.\ 2019b, \mnras, 482, 3373
\bibitem[Fong et al.(2013)]{fong2013} Fong, W., Berger, E., Chornock, R., et al.\ 2013, \apj, 769, 56 
\bibitem[Gao et al.(2016)]{gao2016} Gao, H., Zhang, B., L{\"u}, H.-J.\ 2016, \prd, 93, 044065 
\bibitem[Gao \& Fan(2006)]{gao2006} Gao, W.-H., \& Fan, Y.-Z.\ 2006, \cjaa, 6, 513

\bibitem[Hotokezaka et al.(2013)]{hotokezaka2013} Hotokezaka, K., Kiuchi, K., Kyutoku, K., et al.\ 2013, \prd, 87, 024001 
\bibitem[Just et al.(2015)]{just2015} Just, O., Bauswein, A., Ardevol Pulpillo, R., et al.\ 2015, \mnras, 448, 541
\bibitem[Lattimer, \& Prakash(2010)]{lattimerprakash2010} Lattimer, J.~M., \& Prakash, M.\ 2010, arXiv e-prints, arXiv:1012.3208

\bibitem[Li et al.(2016a)]{li2016a} Li, A., Zhang, B., Zhang, N.-B., et al.\ 2016a, \prd, 94, 083010 
\bibitem[Li et al.(2017)]{li2017} Li, A., Zhu, Z.-Y., \& Zhou, X.\ 2017, \apj, 844, 41 
\bibitem[Li et al.(2016b)]{liye2016} Li, Y., Zhang, B., \& L{\"u}, H.-J.\ 2016b, \apjs, 227, 7 
\bibitem[Li et al.(2018)]{li2018} Li, S.-Z., Liu, L.-D., Yu, Y.-W., et al.\ 2018, \apjl, 861, L12
\bibitem[Luo et al.(2017)]{luo2017} Luo, B., Brandt, W.~N., Xue, Y.~Q., et al.\ 2017, \apjs, 228, 2
\bibitem[L{\"u} et al.(2015)]{lv2015} L{\"u}, H.-J., Zhang, B., Lei, W.-H., Li, Y., \& Lasky, P.~D.\ 2015, \apj, 805, 89 
\bibitem[L{\"u} et al.(2019)]{lv2019} L{\"u}, H.-J., Yuan, Y., Lan, L., et al.\ 2019, arXiv:1904.06664 
\bibitem[Margalit \& Metzger(2017)]{margalitmetzger2017} Margalit, B., \& Metzger, B.~D.\ 2017, \apjl, 850, L19 
\bibitem[Margalit \& Metzger(2019)]{margalit19} -----.\ 2019, \apjl, 880, L15 
\bibitem[Metzger(2017)]{metzger2017} Metzger, B.~D.\ 2017, arXiv:1710.05931 
\bibitem[Metzger \& Piro(2014)]{metzgerpiro2014} Metzger, B.~D., \& Piro, A.~L.\ 2014, \mnras, 439, 3916 
\bibitem[Metzger et al.(2008)]{metzger2008} Metzger, B.~D., Quataert, E., \& Thompson, T.~A.\ 2008, \mnras, 385, 1455
\bibitem[Mosleh et al.(2012)]{mosleh2012} Mosleh, M., Williams, R.~J., Franx, M., et al.\ 2012, \apjl, 756, L12
\bibitem[Peng et al.(2019)]{peng2019} Peng, Z.-K., Yang, Y.-S., Shen, R.-F., et al.\ 2019, \apjl, 884, L34
\bibitem[Piro et al.(2019)]{piro2019} Piro, L., Troja, E., Zhang, B., et al. \ 2019, \mnras, 483, 1912
\bibitem[Pooley et al.(2018)]{pooley2018} Pooley, D., Kumar, P., Wheeler, J.~C., \& Grossan, B. \ 2018, \apj, 859, L23
\bibitem[Rezzolla \& Kumar(2015)]{rezzolla2015} Rezzolla, L., \& Kumar, P.\ 2015, \apj, 802, 95

\bibitem[Rezzolla et al.(2018)]{rezzolla2018} Rezzolla, L., Most, E.~R., \& Weih, L.~R.\ 2018, \apjl, 852, L25 
\bibitem[Rowlinson et al.(2010)]{rowlinson2010} Rowlinson, A., O'Brien, P.~T., Tanvir, N.~R., et al.\ 2010, \mnras, 409, 531 
\bibitem[Rowlinson et al.(2013)]{rowlinson2013} Rowlinson, A., O'Brien, P.~T., Metzger, B.~D., Tanvir, N.~R., \& Levan, A.~J.\ 2013, \mnras, 430, 1061 
\bibitem[Ruiz et al.(2018)]{ruiz2018} Ruiz, M., Shapiro, S.~L., \& Tsokaros, A.\ 2018, \prd, 97, 021501
\bibitem[Shapiro \& Teukolsky(1983)]{shapiroteukolsky1983} Shapiro, S.~L., \& Teukolsky, S.~A.\ 1983, Research supported by the National Science Foundation.~New York, Wiley-Interscience, 1983, 663 p., 
\bibitem[Shibata et al.(2019)]{shibata19} Shibata, M., Zhou, E., Kiuchi, K., \& Fujibayashi, S. 2019, PRD, 100, 023015
\bibitem[Siegel \& Ciolfi(2016a)]{siegel2016a} Siegel, D.~M., \& Ciolfi, R.\ 2016a, \apj, 819, 14 
\bibitem[Siegel \& Ciolfi(2016b)]{siegel2016b} Siegel, D.~M., \& Ciolfi, R.\ 2016b, \apj, 819, 15 
\bibitem[Siegel, \& Metzger(2017)]{siegel2017} Siegel, D.~M., \& Metzger, B.~D.\ 2017, \prl, 119, 231102
\bibitem[Sun et al.(2015)]{sun2015} Sun, H., Zhang, B., \& Li, Z.\ 2015, \apj, 812, 33 
\bibitem[Sun et al.(2017)]{sun2017} Sun, H., Zhang, B., \& Gao, H.\ 2017, \apj, 835, 7 
\bibitem[Woods et al.(1981)]{woods1981} Woods, D.~T., Shull, J.~M., \& Sarazin, C.~L.\ 1981, \apj, 249, 399 

\bibitem[Xiao et al.(2019)]{xiao2019} Xiao, D., Zhang, B.-B., \& Dai, Z.-G.\ 2019, \apjl, 879, L7

\bibitem[Xue et al.(2019)]{xue2019} Xue, Y.~Q., Zheng, X.~C., Li, Y., et al.\ 2019, \nat, 568, 198 
\bibitem[Yang et al.(2019)]{yang2019} Yang, G., Brandt, W.~N., Zhu, S.~F., et al.\ 2019, \mnras, 487, 4721
\bibitem[Yu et al.(2018)]{yu2018} Yu, Y.-W., Liu, L.-D., \& Dai, Z.-G.\ 2018, \apj, 861, 114 
\bibitem[Yu et al.(2013)]{yu2013} Yu, Y.-W., Zhang, B., \& Gao, H.\ 2013, \apjl, 776, L40 
\bibitem[Yuan et al.(2016)]{yuan2016} Yuan, W., Amati, L., Cannizzo, J.~K., et al.\ 2016, \ssr, 202, 235
\bibitem[Zhang(2013)]{zhang2013} Zhang, B.\ 2013, \apjl, 763, L22 
\bibitem[Zhang, \& M{\'e}sz{\'a}ros(2001)]{zhangmeszaros2001} Zhang, B., \& M{\'e}sz{\'a}ros, P.\ 2001, \apjl, 552, L35
\bibitem[Zhang et al.(2018)]{zhang2018} Zhang, B.-B., Zhang, B., Sun, H., et al.\ 2018, Nature Communications, 9, 447 
\bibitem[Zheng et al.(2017)]{zheng2017} Zheng, X.~C., Xue, Y.~Q., Brandt, W.~N., et al.\ 2017, \apj, 849, 127

\end{thebibliography}
\end{document}